\documentclass[sn-apa]{sn-jnl}

\usepackage{graphicx}%
\usepackage{multirow}%
\usepackage{amsmath,amssymb,amsfonts}%
\usepackage{amsthm}%
\usepackage{mathrsfs}%
\usepackage[title]{appendix}%
\usepackage[x11names, table]{xcolor}%
\usepackage{textcomp}%
\usepackage{manyfoot}%
\usepackage{booktabs}%
\usepackage{algorithm}%
\usepackage{algorithmicx}%
\usepackage{algpseudocode}%
\usepackage{listings}%
\usepackage{makecell}
\newcommand{\pkg}[1]{{\normalfont\fontseries{b}\selectfont #1}}
\let\proglang=\textsf
\let\code=\texttt
\usepackage{mdframed}
\usepackage{threeparttable}
\usepackage{multicol}
\usepackage{colortbl} 
\usepackage{soul}

\begin{document}

\title[Article Title]{Fine-grained classification of journal articles based on multiple layers of information through similarity network fusion. The case of the \textit{Cambridge Journal of Economics}}


\author*[1]{\fnm{Alberto} \sur{Baccini}}\email{alberto.baccini@unisi.it}

\author[2]{\fnm{Federica} \sur{Baccini}}

\author[1]{\fnm{Lucio} \sur{Barabesi}}

\author[1]{\fnm{Martina} \sur{Cioni}}

\author[3]{\fnm{Eugenio} \sur{Petrovich}}

\author[1]{\fnm{Daria} \sur{Pignalosa}}

\affil[1]{\orgdiv{Dipartimento di Economia Politica e Statistica}, \orgname{Università degli Studi di Siena}, \orgaddress{\street{Piazza San Francesco, 7}, \city{Siena}, \postcode{53100}, \country{Italy}}}

\affil[2]{\orgdiv{Dipartimento di Ingegneria Informatica, Automatica e Gestionale ``Antonio Ruberti"}, \orgname{Università degli Studi di Roma ``La Sapienza"}, \city{Roma}, \country{Italy}}

\affil[3]{\orgdiv{Dipartimento di Filosofia e Scienze dell'Educazione}, \orgname{Università degli Studi di Torino}, \city{Torino}, \country{Italy}}


\abstract{In order to explore the suitability of a fine-grained classification of journal articles by exploiting multiple sources of information, articles are organized in a two-layer multiplex. The first layer conveys similarities based on the full-text of articles, and the second similarities based on cited references. The information of the two layers are only weakly associated. The Similarity Network Fusion process is adopted to combine the two layers into a new single-layer network. A clustering algorithm is applied to the fused network and the classification of articles is obtained. In order to evaluate its coherence, this classification is compared with the ones obtained by applying the same algorithm to each of two layers. Moreover, the classification obtained for the fused network is also compared with the classifications obtained when the layers of information are integrated using different methods available in literature. In the case of the \textit{Cambridge Journal of Economics}, Similarity Network Fusion appears to be the best option. Moreover, the achieved classification appears to be fine-grained enough to represent the extreme heterogeneity characterizing the contributions published in the journal.}




\maketitle
\bmhead{Acknowledgments}{We thank Alessandra Durio who contributed to the work by doing all the processing for the construction of the similarity matrices based on bags of words and topic modeling. We also thank two anonymous referees for their insightful comments that enabled substantial improvement of the article. This article is available as preprint at \url{https://arxiv.org/pdf/2305.00026.pdf}}

\section{Introduction}\label{sec1}

The classification of scientific papers is a complex task accomplished either by experts or by using suitable algorithms. While expert classification is based on a thorough understanding of the metadata, content, and context of the papers to be classified, algorithmic classification usually relies on a single feature of the papers, such as content, keywords, or references/citations.
For example, algorithmic classifications based on content use distant reading techniques to reveal groups of papers that share similar topics. Alternatively, algorithmic classifications relying on citations use citation relationships among papers to individuate clusters of papers with either overlapping bibliographies, as in bibliographic coupling, or that are frequently cited in the literature, as in co-citation. These clusters can be mapped, in turn, to research fields or subdisciplines. A stream of literature has concentrated on hybrid methods: the classification of articles is conducted after integrating citation- and content-based information with suitable quantitative techniques.

This article intends to contribute to the literature on hybrid methods by exploring the suitability of Similarity Network Fusion (SNF) \citep{wang2014,baccini2022graph,Baccinietal2022,Baccini_2023} for integrating citation- and content-based information to achieve a fine grained-classification of papers. SNF is able to merge different layers of information about articles when they are organized as a multiplex network. 

The full-text and the cited references of articles are considered and similarities among articles are organized in a two-layer multiplex. The first layer conveys similarities based on full-text features, specifically, word frequencies and topics extracted using the Latent Dirichlet Allocation technique \citep{LDA}; the second layer conveys similarities based on cited references, specifically, bibliographic coupling similarities. SNF is used to synthesize the information available in the two separate layers into a new single-layer network, where the previous similarities are properly combined and fused. The fusion process is unsupervised and leverages the structural properties of each layer. The contribution of each layer of information to the determination of the structure of the resulting network is measured using suitable statistical indexes.  
To obtain a classification of papers, a clustering algorithm is applied to the fused network. This classification is then compared with the classification obtained by applying the same algorithm to each of the starting layers. Hence, it is possible to evaluate the coherence between the classification based on all the information available, and the classifications obtained by exploiting only one kind of information at a time. Finally, classifications achieved after the application of SNF are compared with those obtained when the layers of information are integrated using different hybrid methods proposed in literature.

The case study for testing the SNF method covers the set of papers published in the \textit{Cambridge Journal of Economics} during 1985-2013. The choice of the \textit{Cambridge Journal of Economics} is particularly challenging for testing the capacity of the methodology proposed here to generate a meaningful and interpretable fine-grained classification of articles. 

The \textit{Cambridge Journal of Economics} is one of the leading non-mainstream economics journals. It was founded in the 1970s with the main purpose of providing a forum for post-Keynesian, Marxist, and Sraffian scholars, but hosts contributions from all other schools of heterodox thought, such as those in the institutionalist and evolutionary traditions \cite{Saith}. Precisely because its distinguishing character lies in the plurality of approaches, the journal is not dedicated to a single field of economic analysis, but covers a multiplicity of topics ranging from microeconomics to macroeconomics and economic policy, with contributions of both theoretical and applied nature. The papers published in the \textit{Cambridge Journal of Economics} are thus characterized by an extreme heterogeneity both from the viewpoint of their analytical approach and from the viewpoint of the topics covered. Hence the task of classifying such papers is particularly difficult since groups of papers may be defined not only in terms of different topics but also of approaches to the same topic. 

The paper is organized as follows. In Section 2 a short review of relevant literature is provided. Section 3 describes the data. Section 4 describes the methods and the workflow of the exploratory analysis. Results are presented in Section 5. Section 6 discusses and interprets the results. Section 7 concludes by suggesting further steps for the present line of research.

\section{A short literature review}\label{sec2}

According to \cite{glanzel_2003} ``the classification of science into a disciplinary structure is at least as old as science itself''. From Aristotle to Medieval logicians up to Nineteenth-century positivists, philosophers and scientists have proposed numerous classificatory schemes for organizing human knowledge \citep{fisher_classification_1990}. 
In the Twentieth century, many concurring general classification schemes have become established, such as the Dewey classification, the OECD's fields of science, and Web of Science (WoS) or Scopus categories (for a short review see the entry ``research fields'' in \citealp{RN1}). These general schemes, however, ``are too broad to adequately capture the more complex, fine-grained cognitive reality'' \citep{Eykens}. Hence, uncountable attempts have been developed for classifying disciplines at the desired fine-grained level.

From a theoretical point of view, the delineation of scientific fields consists in partitioning the objects of the analysis into groups by using some classification technique. These techniques are divided by \cite{Zitt2019} into three distinct groups: (i) ready-made or institutional classifications of science which originate from scientists or librarians and do not entail the use of bibliometrics; each ``artifact'' is classified by experts who assign it the correct label after a thorough evaluation based on its contents, context, and metadata; (ii) ex-post classifications where experts attach disciplinary or sub-disciplinary labels to clusters of ``artifacts'' obtained by network analysis techniques applied to relevant bibliometric or scientometric data; (iii) classifications that rely on highly supervised schemes functional to efficient information retrieval.

Most of the literature considers articles as the basic unit of scientific fields, and hence as the target of the classification.
Similarities between pairs of papers have been defined in terms of contents by considering title and abstracts \citep{Boyack_2017}, keywords or text \citep{Ahlgren_2009}; in terms of citation relations such as direct citations \citep{sjogarde_2018, sjogarde_2020} or co-citations \citep{small1973} or bibliographic coupling \citep{kessler_1965}; or, finally, in terms of social attributes such as authorship \citep{ni_2013}. Many works aim to compare scientific field classifications emerging from the use of different definitions of similarity relations among articles \citep{Boyack_Klavans_2010, Klavans_2017, Kleminski_2020}. \citet{sjogarde_2018} compare the choice of different parameters of modularity optimization for obtaining a fine-grained classification of disciplines.
 
By and large, the big part of the literature devoted to scientific fields delineation uses an approach based on a single-layer network, by exploiting only one type of information for classification or for the definition of a science map \citep{Zitt2019,Petrovich2020}. \citet{Zitt2019} labeled as ``hybridization'' or ``multinetwork approaches'' those contributions that try to combine different layers of information for scientific fields delineation and classification purposes. Recent reviews are \citet{Zitt2019} and \citet{Boyack2020}. Usually, the integration of multiple layers of information requires the heavy intervention of researchers for their integration. The most-used strategy is to combine the two layers of information about relatedness of papers with  a weighted linear combination, as reviewed by \citet{Boyack2020}. The choice of weights is not an easy task, and different choices may produce different results.
\citet{Glanzel2011,glanzel_thijs}, by developing an idea proposed in \cite{janssens}, combined information obtained by bibliographic coupling and textual similarities as the cosine of a weighted linear combination of the arccosine trasformation of similarities, and studied the different results obtained by the adoption of different weights. 

The SNF technique adopted in this paper for integrating different layers of information about papers has been applied in informetrics for the first time to the classification of scholarly journals. Specifically, \cite{Baccinietal2022} used the similarity network fusion technique for classifying journals by considering a three-layer network: the first layer is generated by bibliograhic coupling among journals, the second considers the crossed presence of the same authors contributing to different journals, the third the crossed presence of the same scholars in the editorial boards of different journals. Differently from the other techniques of integration recalled above, the SNF is completely unsupervised and does not require assumptions about the data or about the weights to assign to different information during the integration process. Furthermore, as will be illustrated later on, it is possible to measure \textit{ex-post} the contribution of each layer of information to the structure of the fused final network \citep{Baccini_2023}.

As for the case-study developed here, it regards economics. The classifications of economics adopted by Web of Science and Scopus are rough:  Scopus classified articles as Economics, Econometrics or Finance; WoS as Economics, Business and Economics, and Business and Finance. At the opposite the JEL codes, i.e.\ the expert classification usually adopted by economists, is hierarchically organized in 20 macro fields and hundreds fine grained codes \citep{Cherrier_2017}.
Only in the last decade have economists developed interest in quantitative methods as tools to improve their understanding of the structure and evolution of their discipline \citep{edwards}. A few contributions have employed citation information, both in the form of bibliographic coupling and co-citation analysis for classification purposes. For instance, \citet{Claveau} use bibliographic coupling to measure the cognitive similarity between articles in a corpus of over 400,000 documents retrieved from the Web of Science database and construct a dynamic network analysis that leads to identifying families of research fields and their evolution over time. They are thus able to identify the emergence and decline of subfields within economics since the late 1950s and reconstruct the history of specialties in the discipline. \citet{Truc_2021} rely on co-citation between articles published in the two main economic methodology journals over the past three decades in order to appraise the standard interpretation of the developments in the field. They generate three co-citation networks, one for each decade under consideration, and, by observing continuities and changes across the networks, they assess the main historical trends put forth in the existing interpretive literature.
\cite{Ambrosino_2018} applied LDA topic modeling technique to construct a map of economics over time and detect key developments in the structure of the discipline. 
An attempt at combining citation and content information was carried out by \citet{Garcia_2023}. They analyze the developments od the small field of consumption modeling over forty years by constructing co-citation networks and combining them with semantic evidence in the form of the most frequent strings of words used in the abstracts of co-citing articles. 

\section{Data}\label{sec3}
Cited references data and textual data were retrieved from Web of Science (WoS) and JSTOR databases, respectively. JSTOR archival journal collection includes more than 2,800 academic journals across the humanities, social sciences, and natural sciences from 1,200 publishers from 57 countries.
The time span of the analysis was determined by the data availability in the two databases: WoS started recording the \textit{Cambridge Journal of Economics} in 1985, whereas JSTOR does not provide access to its most recent issues because of JSTOR policies. At the time of data retrieval (2019), the last complete available year was 2013. Thus, the time span was set from 1985 to 2013. 

WoS data, including the cited references of the records, were retrieved from WoS web platform, whereas the $n$-grams used in the topic modeling were retrieved from JSTOR Data for Research platform. In both cases, the query was based on the title of the target journal. The records in the two datasets were then matched using the volume, number, and name of the first authors. Any record without cited references or appearing in only one of the datasets was excluded, so that the final dataset included 1,344 records. Note that all types of documents published by the \textit{Cambridge Journal of Economics}, not only research articles, were included.



To improve the reliability of citation analysis, cited references were cleaned using the CRExplorer software \citep{thor_introducing_2016}. CRExplorer individuates and merges variants of the same reference through an algorithm based on string similarity. The process was humanly supervised to avoid wrong merging and individuate further variants to unify. Special attention was reserved for books and historical references, which are very common in this journal, in order to merge all of their possible variants. Of the 45,611 distinct cited references appearing in the raw dataset, 42,072 distinct cited references remained after this consolidation process (-7.8\%). A significant number of cited references (84.7\%) collects only 1 citation in the entire dataset. 

The 1,344 papers retrieved from JSTOR contained a total of 7,540,085 non-distinct words.
This corpus of textual data was prepared in a suitable way for the topic modeling analysis using several \proglang{Python} scripts \citep{bird2009}. Pre-processing included the normalization of words with anomalous characters such as numbers or accents, the removal of too short and too long words, which stem from errors in the original documents, and the filtering out of stop-words. To scale down the size of the vocabulary, moreover, words were reduced to their root form (stems) through stemming, which was implemented with the \pkg{nltk} library. Lastly, rare stems that occurred in less than three papers in the corpus and common stems that occurred in more than 95\% of the documents were removed. The processing of the textual data resulted thus in a vocabulary of 79,262 distinct stems distributed among 1,344 documents.

\section{Workflow and methods of the exploratory analysis}\label{sec4}

The basic building blocks of the analysis are two different ways of representing similarity relations among articles, based respectively on cited references and on contents: words and topics extracted with LDA. Thus, the first step of the analysis consists in defining the layers of the similarity network (i.e.\ the multiplex). 

As to the layer based on cited references, each paper is characterized by the set of its cited references, and the similarity between two articles is computed by considering their common cited references. The basic information is organized in a bipartite network, where the first set of nodes contains citing articles and the second set of nodes contains the cited references. Edges link each citing paper to its cited references. Similarities between each pair of citing articles are then computed by using the classical Jaccard similarity coefficient \citep{Jaccard} and organized in a square similarity matrix, representing the first layer of the multiplex. More precisely, if $A_i$ and $A_j$ represent the sets of cited references of the $i$-th and $j$-th paper, the Jaccard coefficient is defined as
\begin{equation}
J_{ij}=\frac{| A_i \cap A_j |}{| A_i \cup A_j |}\ ,
\end{equation}
where $\mid\cdot\mid$ denotes the cardinality of a set. It is apparent that $0\le J_{ij}\le 1$. Hence, the similarity between two articles is proportional to the number of references cited by both papers: when two papers cite exactly the same set of references, i.e.\ when $A_i=A_j$, the maximum similarity $J_{ij}=1$ occurs. In contrast, the minimum similarity $J_{ij}=0$ is achieved when two papers have no common references, i.e.\ when $A_i\cap A_j=\emptyset$. Similarities between each pair of articles based on Jaccard index are collected in a square similarity matrix, representing the first layer of a multiplex.

As to the layer of information based on contents, the basic idea is to measure similarity between articles in terms of similarity of contents. The adoption of a distant reading perspective over the full-text of articles is straightforward and it is here implemented in two different ways. The first one is very simple and does not require theoretical assumptions: each paper is characterized by its ``Bags of Words'' (hereafter BoW), i.e. the frequent distribution of lexical items used in it.
Thus, the similarity between each pair of papers is computed by considering their common words. For implementing this approach, the basic information is organized in a bipartite network where the first set of nodes contains articles and the second set contains the words used. Edges link each article to its words with a weight proportional to the word frequency in the article. Under this setting, it is presumed that the style of an article is mainly revealed through the choice of words, and in particular by the very frequent words in the whole corpus. For more details, see the survey provided by \citet[Chapter 3]{Savoy}. The word types do not have a precise meaning, and they induce a stylistic description, which is independent of the topics of the underlying article. To be more explicit, let us assume that there exist $L$ most frequent word types in the selected articles, without taking punctuation marks or numbers into consideration. In addition, let $p_{il}$ be the relative frequency of the $l$-th word in the corpus for the $i$-th paper. Thus, the matching between a pair of papers may be computed by using a similarity concept based on the total variation measure, and it is given by
\begin{equation}\label{eq::total_var}
V_{ij}=1-\frac{1}{2} \sum_{l=1}^{L}|p_{il}-p_{jl}|\ . 
\end{equation}
In fact, the normalized total variation measure between the $i$-th and $j$-th article is nothing else than $(1-V_{ij})$. The total variation measure is usually (and very naturally) adopted for comparing two categorical distributions (see e.g.\ \citealp{agresti}). In turn, we have that $0\le V_{ij}\le 1$. In particular, if two articles have the same relative frequency of words in the corpus, the maximum similarity $V_{ij}=1$ occurs. On the other hand, the minimum similarity $V_{ij}=0$ is reached when the two articles adopt completely different words. In the stylometric framework, it is worth noting that $(1-V_{ij})$ is the Labbé’s inter-textual distance (for more details see \citealp{Savoy}). Similarities between each pair of articles based on BoW are organized in a square similarity matrix, representing the second layer of a multiplex.

The second approach is based on topic modeling. Under this framework, a pre-defined number of ``topics'', say $K$, is considered, and the Latent Dirichlet Allocation (LDA) method is carried out on the full-text of the whole set of articles \citep{LDA}. A detailed account of LDA is given by \citet[section 7.3]{Savoy}. LDA produces $K$ lists of words and the distributions of the lists in the whole corpus. Each word list and the corresponding distribution is defined as a topic in LDA. Hence, each paper is associated with a topic distribution. In such a case, we obtain a bipartite network where the first set of nodes contains articles, and the second set contains topics. Edges link each paper to the topics with a weight proportional to the topic frequency. Similarities between pairs of papers can be computed by using the similarity index in \eqref{eq::total_var}, where $p_{il}$ is the relative frequency of the $l$-th topic for the $i$-th paper, and $L$ is replaced by $K$. Also in this approach, similarities between each pair of citing articles are organized in a square similarity matrix, representing a layer of a multiplex.

As anticipated, LDA requires the topic number $K$ as a smoothing parameter to be defined, and no obvious selection rule can be generally given. A possible empirical strategy is to define different numbers of topics to construct different similarity matrices and test the stability of the obtained results. 

\subsection{Comparing similarity matrices}\label{sec5}

The subsequent step of the exploratory analysis consists in computing the association among the structures of these similarity matrices for verifying the coherence of the information contained in them. 

In order to compare the dependence between the similarity matrices, the generalized distance correlation $R_d$ suggested by \cite{szekely2007} is adopted. Its interpretation is similar to the squared Pearson correlation coefficient: $R_d$ is defined in the interval $[0,1]$; values close to zero indicate no or very weak association; larger values indicate a stronger association, which is perfect for $R_d=1$, and similar considerations hold for $\sqrt{R_d}$ (for more details, see \citealp{RN23}). A high value of distance correlation indicates that two different similarity matrices, based on cited references, and on words and topics, convey the same information. On the contrary, a low value of distance correlation indicates that they convey different information.

\subsection{Detecting and comparing communities of articles}\label{sec6}

In the third step of exploratory analysis, the similarity networks are partitioned into communities or clusters of articles by using Louvain algorithm based on modularity \citep{RN37}. The clusters of papers obtained are then compared by using suitable statistical techniques, namely by computing Cramer's V \citep{Cramer}. 

Results of this step will indicate if the clusters obtained in different networks are coherent or not, and may or may not reinforce the ones obtained after the correlation analysis. Consider for instance the simplest case in which clusters obtained in the networks are highly similar, and distance correlations indicate that similarity networks, based on cited references and on words and topics, convey the same information. It can be concluded that cited references and content-based networks convey the same information and produce similar classifications of articles, i.e. very similar clusters.
Conversely, if the distance correlation is low and the obtained clusters are dissimilar, it could be useful to consider all the available information. 

\subsection{Applying similarity network fusion}

The application of SNF technique permits to synthesize in a single layer the information contained in multiple layers \citep{wang2,wang2014, Baccinietal2022, baccini2022graph}.  
In the present case, the SNF is realized by considering the two layers of the multiplex formed by cited references and by one of the matrices based on topics, or, alternatively, on bags of words. 

SNF integrates into a unique similarity matrix the pair of original matrices by means of the Cross Diffusion Process (CDP) \citep{wang2,wang2014}. CDP is an unsupervised iterative procedure that reinforces very strong links present in the single layers, and those that are common to all the layers. 
As a result, SNF generates a new network where nodes are articles and edges are weighted according to the new similarity values obtained through CDP. In fact, the iterative procedure enriches the information of a single layer with that coming from the other layer. SNF maintains densely connected groups of articles and reinforces their links; at the same time, SNF makes more visible low weighted links that are present in both the layers, as they may represent stable relationships among groups of articles. The generalized distance correlation can be used for evaluating how much of the information contained in the two single layers is reported also in the fused networks. 

Furthermore, the partial distance correlation $R_d^*$ proposed by \cite{székely2014} can be used for analyzing the contribution of each layer to the similarity structure of the fused networks, as in \citet{RN40}. Partial distance correlation measures 
the degree of association between the similarity matrix of the fused network and a layer, by removing the effect of the second layer. 

As a final exploratory step, articles are classified by using the Louvain algorithm in the fused networks. The stability of classifications among different fused networks is evaluated by Cramer's V. The interpretation of an expert will indicate if the classifications obtained in the fused networks provide a fine-grained classification of articles that is more satisfactory than the classifications obtained by considering cited references and words and topics separately.

\subsection{Comparison with other hybrid models}
The results of SNF are compared with the ones reached by adopting other hybrid models, i.e. other techniques of integrating information organized in two matrices that are recalled in the literature review.
\citet{Boyack2020} proposed to integrate two layers of information by means of a convex combination of matrices. 
Let $\mathbf{S}_1=(s_{1,ij})$ and $\mathbf{S}_2=(s_{2,ij})$ be two similarity matrices of order $(n\times n)$. The average of these similarity matrices consists in considering the convex combination $\mathbf{S}=(s_{ij})$
given by
\begin{equation*}
\mathbf{S}=\alpha \mathbf{S}_1+(1-\alpha) \mathbf{S}_2,
\end{equation*}
where $\alpha\in(0,1)$ is given by
\begin{equation*}
\alpha=\frac{T_2}{T_1+T_2}
\end{equation*}
with
\begin{equation*}
T_1=\sum_{i=1}^n\sum_{j=1}^ns_{1,ij},T_2=\sum_{i=1}^n\sum_{j=1}^ns_{2,ij}.
\end{equation*}
For practical purposes, \citet{Boyack2020} also proposed to use three further different sets of predefined weights, by attributing the lowest weight to the matrix with the highest value between $T_1$ and $T_2$. It should be remarked that this proposal is the Fr\a'echet mean of the input matrices, obtained by assuming the choice of the Frobenius metric, as discussed by \citet{Baccini_2023}, where other advanced methods for selecting the weights are also proposed.
In addition, \citet{Glanzel2011} proposed to integrate two similarity matrices, by attributing weights to their transformations, according to 
\begin{equation*}
s_{ij}=\cos(w\arccos(s_{1,ij})+(1-w)\arccos(s_{2,ij})),
\end{equation*}
where $w\in(0,1)$. 
For practical purposes, equal weights are often chosen in this case, i.e.\ $w=1/2$ is selected.

\section{Results}

The first step of analysis consists in constructing the similarity matrices. They are symmetric $(1,344\times1,344)$ matrices containing similarities between each pair of CJE articles. A total of eights similarity matrices were constructed: (i) a matrix where similarities are computed on cited references (``Cited References"); (ii) a similarity matrix based on bags of words (``Bags of Words"). As anticipated, LDA requires the choice of the topic number $K$; the empirical strategy adopted here was to assume $K=5,10,15,20,25,30$, and construct (iii) six similarity matrices, each one referring to the selected topic number K (``Topics\_5", \dots, ``Topics\_30").  

\subsection{Comparing similarity matrices} 

As anticipated the comparison between similarity matrices is conducted through distance correlations; they were computed in the \proglang{R}-computing environment \citep{RN13} by using the \code{dcor} function in the package \pkg{energy}. The generalized distance correlation between matrices based on topics reported in Table \ref{dcor_topic} allows to check whether the information obtained changes by using different topic numbers $K$. Results indicate that when a $K$ higher than 5 is chosen, the information obtained is substantially similar, with values of generalized distance correlations generally higher than 0.9. The similarity matrix obtained by setting $K=5$ has, by contrast, the lowest association with the other matrices.

The generalized distance correlation between the matrices based on topics and the similarity matrix based on BoW allows, on the other hand, to check if the use of BoW and of topic modeling produce similar information in terms of similarities between articles. The generalized distance correlations, also reported in Table \ref{dcor_topic}, are stable around 0.71 regardless of the number of topics. This result can be interpreted as indicating that the use of topics modelling does not entail a relevant loss of information with respect to the use of BoW. 

\begin{sidewaystable}
\scriptsize
\caption{Generalized distance correlation between article similarity matrices.}
    \centering
    \begin{tabular*}{\textwidth}{@{\extracolsep\fill}ccccccccc}
    \toprule%
            dCor & Topics\_5 & Topics\_10 & Topics\_15 & Topics\_20 & Topics\_25 & Topics\_30 & Bags of words & Cited References \\ \hline
        Topics\_5 & 1 & 0.830 & 0.828 & 0.828 & 0.810 & 0.799 & 0.712 & 0.408\\
        Topics\_10 & ~ & 1 & 0.929 & 0.905 & 0.889 & 0.8892 & 0.712 & 0.464 \\ 
        Topics\_15 & ~ & ~ & 1 & 0.950 & 0.934 & 0.926 & 0.707 & 0.492 \\ 
        Topics\_20 & ~ & ~ & ~ & 1 & 0.974 & 0.961 & 0.712 & 0.501 \\ 
        Topics\_25 & ~ & ~ & ~ & ~ & 1 & 0.980 & 0.710 & 0.514 \\ 
        Topics\_30 & ~ & ~ & ~ & ~ & ~ & 1 & 0.706 & 0.519 \\ 
        Bags of Words & ~ & ~ & ~ & ~ & ~ & ~ & 1 & 0.455 \\ 
        Cited References  & ~ & ~ & ~ & ~ & ~ & ~ & ~ & 1 \\ \bottomrule
    \end{tabular*}
    \label{dcor_topic}
\end{sidewaystable}

Finally, the generalized distance correlations between the article similarity matrix based on cited references and, respectively, the matrix based on BoW and the matrices based on topics indicate if the information contained is associated or not. In case of a very high distance correlation between matrices, one could argue that the choice of one or the other matrix is not relevant, as the information conveyed by both matrices is highly associated. On the opposite, a very low distance correlation could indicate that the matrix based on cited references and matrices based on BoW or topics convey different information and then both kinds of information should be considered. 

The last column of Table \ref{dcor_topic} shows that the generalized distance correlation between the matrix based on cited references and the matrices based on topics has a value that tends to a slight growth as the number of topics increases. Also the generalized distance correlation between the matrix based on cited references and the matrix based on BoW has an intermediate value. The ambiguity of all these results suggests, however, that the information conveyed by cited references and contents is not highly associated and, as a consequence, both information sources should be taken into consideration.

\subsection{Detecting and comparing communities of articles} 

The eight matrices are then used to construct as many different classifications of items, by using the Louvain algorithm. Table \ref{clusters} reports the number of clusters and the corresponding values of modularity.

\begin{table}[!ht]
\scriptsize
\caption{Clusters of articles obtained through the Louvain algorithm applied to article similarity matrices based on cited references, topics and bags of words.}
    \centering
    \begin{tabular}{ccc}
    \toprule
&n. of clusters & Modularity \\ \hline
Topics\_5&5&0.351\\
Topics\_10&4&0.315\\
Topics\_15&5&0.310\\
Topics\_20&5&0.321\\
Topics\_25&5&0.324\\
Topics\_30&5&0.324\\
BoW&3&0.039\\
Cited References&51&0.410\\ \bottomrule
    \end{tabular}
    \label{clusters}
\end{table}

The Louvain algorithm applied to the similarity matrix based on cited references produces 51 clusters (modularity $0.41$). Only 6 clusters have more than 100 articles for a total of 1,278 articles ($95.1\%$ of the total number of articles), 4 contain less than 8 articles (for a total of 22 articles, $1.6\%$), and the remaining 41 clusters are composed mainly by isolated articles (for a total of 44 articles $3.3\%$). When the Louvain algorithm is applied to similarity matrices based on topics, for any number of topics, it produces 5 clusters with the only exception of the case of Topic\_10 which results in 4 clusters. The modularity values are very similar for any number of topics. Finally, when the similarity of articles in terms of BoW is considered, the Louvain algorithm produces 3 clusters with very low modularity, indicating that it is unable to neatly separate clusters of articles. 

\begin{sidewaystable}
\scriptsize
\caption{Values of Cramer's V for measuring the association between clusters obtained from article similarity matrices based on cited references, topics and bags of words.}
    \centering
    \begin{tabular*}{\textwidth}{@{\extracolsep\fill}ccccccccc}
    \toprule
         & Topics\_5 & Topics\_10 & Topics\_15 & Topics\_20 & Topics\_25 & Topics\_30 & BoW & CR \\ \hline
Topics\_5&1&0.722&0.676&0.699&0.723&0.728&0.806&0.639\\
Topics\_10&&1&0.800&0.777&0.789&0.783&0.798&0.572\\
Topics\_15&&&1&0.826&0.772&0.761&0.798&0.521\\
Topics\_20&&&&1&0.842&0.801&0.792&0.645\\
Topics\_25&&&&&1&0.891&0.792&0.535\\
Topics\_30&&&&&&1&0.794&0.538\\
BoW&&&&&&&1&0.512\\
CR&&&&&&&&1\\ \bottomrule
    \end{tabular*}
    \label{cramer}
\end{sidewaystable}

In Table \ref{cramer}, the values of Cramer's V are reported for measuring the associations between classifications obtained for different numbers of topics, bags of words, and cited references. All the classifications obtained by considering topics are highly associated. The classifications for Topic\_15, Topic\_20, and Topic\_25 have particularly high values of Cramer's V. Analogously, all the clusters obtained for different numbers of topics are highly associated with the clusters obtained for bags of words: this provides further support to the observation made above that the use of topics does not entail a loss of information compared to the use of BoW. Finally, the association between clusters based on cited references and the ones based on topics and bags of words shows intermediate values. This is coherent with the result obtained by considering the generalized distance correlations: the information conveyed by the similarity network based on cited references and the ones based on BoW and topics is not highly associated.  

In sum, this step of analysis suggests that article similarity matrices based on topics and on BoW contain entirely similar information and result in overlapping papers classifications. The similarity matrix based on cited references instead appears to convey information different from the matrices based on topics and bags of words, and, accordingly, the classification obtained from it does not overlap with the others.

\subsection{Similarity network fusion} 
SNF is used to integrate the information contained in the similarity network based on cited references and in the similarity matrices based on topics and bags of words.  SNF was realized in the \proglang{R} environment by using the \code{SNF} function in the \pkg{SNFtool} package (R Core Team, 2020). A total of seven fused networks are constructed, corresponding to the fusion of the similarity matrix based on cited references, with each of six similarity matrices based on topics (Fused\_5, \ldots , Fused\_30), and with the matrix based on bags of words (Fused\_BoW). 

The generalized distance correlation can be used for evaluating how much of the information contained in the two single layers is reported also in the fused networks. The generalized distance correlations between the similarity matrix based on cited references and each of the seven fused networks are all very high ($0.957$). The generalized distance correlation between each of the fused networks and the corresponding similarity matrices based on topics and on BoW is stable and low (around $0.397$).  

For verifying if the SNF generates different structures of fused networks according to the number of topics, it is possible to use again the generalized distance correlations among fused matrices. The computed values of generalized distance correlations are all near $1$, indicating that the structures of the fused matrices are identical, regardless of the number of topics chosen or the choice of BoW.

\begin{sidewaystable}
\centering
\scriptsize
\caption{Partial distance correlation between the seven fused networks and the two corresponding layers. Each row reports the values of partial distance correlation between the fused networks and one of the layers, conditioned on the other layer.}
\bigskip
\begin{tabular}{ccccccccc}
\toprule
& & Fused\_5 & Fused\_10 & Fused\_15 & Fused\_20 & Fused\_25 & Fused\_30 & Fused\_BoW \\
\bottomrule
\\
Fused, Topics$|$Cited references & $\sqrt{R_d^*}$ & 0.177&0.208&0.232&0.244&0.254&0.247&0.094 \\ 
\\
Fused, Cited references$|$Topics & $\sqrt{R_d^*}$ & 0.319&0.308&0.307&0.302&0.298&0.297&0.335 \\ 
 \\
\hline		
\end{tabular}
\label{partial_dist}
\end{sidewaystable}

As anticipated, differently from other methods of hybridization, SNF does not require assumptions about the weights to attribute to different layers of information. Nonetheless, the partial distance correlation $R_d^*$ permits to measure the contribution of each of the two similarity matrices to the structure of the resulting fused matrix. 
The partial distance correlations were evaluated in t he \proglang{R}-computing environment \citep{RN13} by using the \code{pdcor} function in the package \pkg{energy}. The computed values of partial distance correlations are reported in Table \ref{partial_dist}. The partial distance correlations indicate that the layer of cited references gives the major contribution to the structure of the fused networks, regardless of the number of topics chosen or the choice of BoW.        

Also in this case, the classification of articles in the seven fused networks is conducted by using the Louvain algorithm. Table \ref{clusters_in_fused} reports the number of resulting clusters and the modularity values. The number of clusters lies between 7 and 9 for fusions based on topics, while it goes up to 11 when the fusion is based on bags of words. The values of modularities are always similar.

\begin{table}[!ht]
\scriptsize
\caption{Clusters of articles obtained through Louvain algorithm applied to fused networks.}
    \centering
    \begin{tabular}{ccc}
    \toprule
&n. of clusters & Modularity \\ \hline
Fused\_5&7&0.427\\
Fused\_10&7&0.426\\
Fused\_15&9&0.430\\
Fused\_20&8&0.432\\
Fused\_25&9&0.432\\
Fused\_30&8&0.430\\
Fused\_BoW&11&0.445\\
\bottomrule
    \end{tabular}
    \label{clusters_in_fused}
\end{table}

The stability of the classifications between all the possible pairs of fused networks is also explored by estimating the Cramer's V, as reported in Table \ref{cramer_fused}. These values indicate that the classifications have a relatively high degree of association that tends to grow with the number of topics used in the fusion. The clustering obtained for the Fused\_BoW matrix has high and stable degrees of association with all the clustering obtained in the other fused matrices. 

\begin{table}[!ht]
\scriptsize
\caption{Values of Cramer's V for measuring the association between classifications of articles in the fused networks.}
    \centering
    \begin{tabular}{cccccccc}
    \toprule
         & Fused\_5 & Fused\_10 & Fused\_15 & Fused\_20 & Fused\_25 & Fused\_30 & Fused\_BoW  \\ \hline
Fused\_5&1&0.624&0.619&0.670&0.717&0.613&0.848\\
Fused\_10&&1&0.661&0.589&0.595&0.585&0.799\\
Fused\_15&&&1&0.696&0.731&0.749&0.712\\
Fused\_20&&&&1&0.730&0.737&0.746\\
Fused\_25&&&&&1&0.644&0.788\\
Fused\_30&&&&&&1&0.704\\
Fused\_BoW&&&&&&&1\\
 \bottomrule
    \end{tabular}
    \label{cramer_fused}
\end{table}

\subsection{Comparison with other hybrid models} 
In Section \ref{sec2} other methods of integrating information organized in similarity matrices are summarily presented. SNF results can be compared with a couple of these methods. For the sake of simplicity, since the structure of the fused networks are identical, the comparison is limited to the case of Fused\_20.
The most recent contribution is \cite{Boyack2020}. They propose a method for computing the weights for integrating the two layers of information that, in the present case, results in a weight of $0.018$ for topics and $0.982$ for cited references. The resulting matrix is refereed to as ``BK". According to their proposal, other matrices are also constructed by assuming respectively equal weights (``BK1"), $0.333$ for topics and $0.667$ for cited references (``BK2"), and $0.2$ for topics and $0.8$ for cited references (``BK3").

The second method considered is the one proposed by \citet{Glanzel2011} reported in Section \ref{sec2}; the obtained matrix is denoted as ``GT".

\begin{table}
\scriptsize
\caption{Generalized distance correlation between similarity matrices obtained by SNF and other methods of integrating information.}
    \centering
    \begin{tabular*}{\textwidth}{@{\extracolsep\fill}ccccccc}
    \toprule%
        dCor & BK & BK1 & BK2 & BK3 & GT & Fused\_20 \\ \hline
        BK & 1 & 0.689 & 0.701 & 0.735 & 0.687 & 0.929 \\
        BK1 & ~ & 1 & 1 & 0.998 & 0.999 & 0.426  \\ 
        BK2 & ~ & ~ & 1 & 0.999 & 0.998 & 0.443 \\ 
        BK3 & ~ & ~ & & 1 & 0.996 & 0.489 \\ 
        GT & ~ & ~ & ~ & & 1 & 0.425 \\ 
        Fused\_20 & ~ & ~ & ~ & ~ & & 1  \\ 
    \hline
    \end{tabular*}
    \label{dcor_boyack}
\end{table}

Also in this case, the generalized distance correlations can be computed to compare the structures of the matrices obtained with different integration methods, as reported in Table \ref{dcor_boyack}. The table highlights that the matrices BK1, BK2, BK3 and GT are substantially identical. Moreover, they show a high distance correlation value with the BK matrix, but a moderate correlation value with the matrix Fused\_20. The fused matrix has instead a high value of correlation with the BK matrix. This can be explained by the fact that the layer of cited references contributes the most to the fused network, and BK is built by reducing to a minimum the weight attributed to the matrix based on topics, as previousely observed.

The search for communities in the matrices obtained by using other methods of integrating information results, as reported in Table \ref{clusters_in_boyack}, in a smaller number of communities and lower modularity values with respect to the Fused\_20 matrix (see Table \ref{clusters_in_fused}).

\begin{table}[!ht]
\scriptsize
\caption{Clusters of articles (Louvain algorithm) in networks obtained by other methods of integrating information.}
    \centering
    \begin{tabular}{ccc}
    \toprule
&n. of clusters & Modularity \\ \hline
BK&4&0.327\\
BK1&6&0.318\\
BK2&5&0.319\\
BK3&5&0.318\\
GT&6&0.330\\
\bottomrule
    \end{tabular}
    \label{clusters_in_boyack}
\end{table}

The computation of Cramer's V reported in Table \ref{cramer_boyack} shows that the classification of articles obtained in BK1, BK2, BK3 and Glanzel have a very high association; the classification obtained in the Fused\_20 network has lower value of association.  

\begin{table}
\scriptsize
\caption{Values of Cramer’s V for measuring the association between classifications of articles in the networks obtained by SNF and other methods of integrating information.}
    \centering
    \begin{tabular*}{\textwidth}{@{\extracolsep\fill}ccccccc}
    \toprule%
            dCor & BK & BK1 & BK2 & BK3 & GT & Fused\_20 \\ \hline
        BK & 1 & 0.778 & 0.723 & 0.790 & 0.808 & 0.731 \\
        BK1 & ~ & 1 & 0.890 & 0.911 & 0.857 & 0.665 \\ 
        BK2 & ~ & ~ & 1 & 0.931 & 0.943 & 0.703 \\ 
        BK3 & ~ & ~ & & 1 & 0.933 & 0.679 \\ 
        GT & ~ & ~ & ~ & & 1 & 0.655 \\ 
        Fused\_20 & ~ & ~ & ~ & ~ & & 1  \\ 
    \bottomrule
    \end{tabular*}
    \label{cramer_boyack}
\end{table}

The comparison of the results obtained with SNF and other methods of integration clearly indicates that they do not coincide. Moreover, the choice of weights is the most delicate passage in both the proposals by \citet{Boyack2020} and by \citet{Glanzel2011}. Choosing different weight systems without studying the structure of the matrices to be integrated, as in BK1, BK2, BK3 and GT, results in very similar weighted matrices and very similar classifications. The choice proposed by \citet{Glanzel2011} does not differentiate the final results from the ones obtained by the simple weighted means proposed by \citet{Boyack2020}. 
In the present application, the more sophisticated method of computing the weights proposed by \citet{Boyack2020} and reported in Section \ref{sec3}, gives rise to a completely unbalanced weighting of the two matrices, that probably seizes the hidden structure of the two input matrices. This BK matrix is highly correlated with the Fused\_20 matrix obtained through SNF, where one of the two input matrices contributed the most to the final structure.
Despite the high correlation between the two matrices, the classifications of articles obtained in them are not strongly associated.

\section{Main characteristics of groups of papers}

In order to test the meaning and the interpretability of the classification obtained in the SNF-built network, an analysis of experts in the field has been carried out.  As anticipated, the papers published in the \textit{Cambridge Journal of Economics} are characterized by an extreme heterogeneity both from the point of view of their analytical approach and of the topics covered. This heterogeneity is reflected in the low value of distance correlation ($0.501$, see Table \ref{dcor_topic}) between the papers' similarity matrices obtained from textual information and from cited references. On the other hand, the classifications obtained for fused networks are highly stable. Hence, in what follows the Fused\_20 network case only is analyzed and discussed in detail. Note that the case of Fused\_20 is the less favorable for illustrating the utility of fusing information instead of relying on only one of the two layers of information. Indeed, the value of Cramer's V for the association between clusters obtained in the network based on cited references and on Topic\_20 is the highest, i.e. the information conveyed by the two networks has the highest level of association with respect to all the other cases. Hence, valuable results in the most challenging case are a good premise to argue about the utility of fusion in less challenging cases, when the association between layers is lower.

\begin{sidewaysfigure}
\centering
     \includegraphics [scale=0.6]{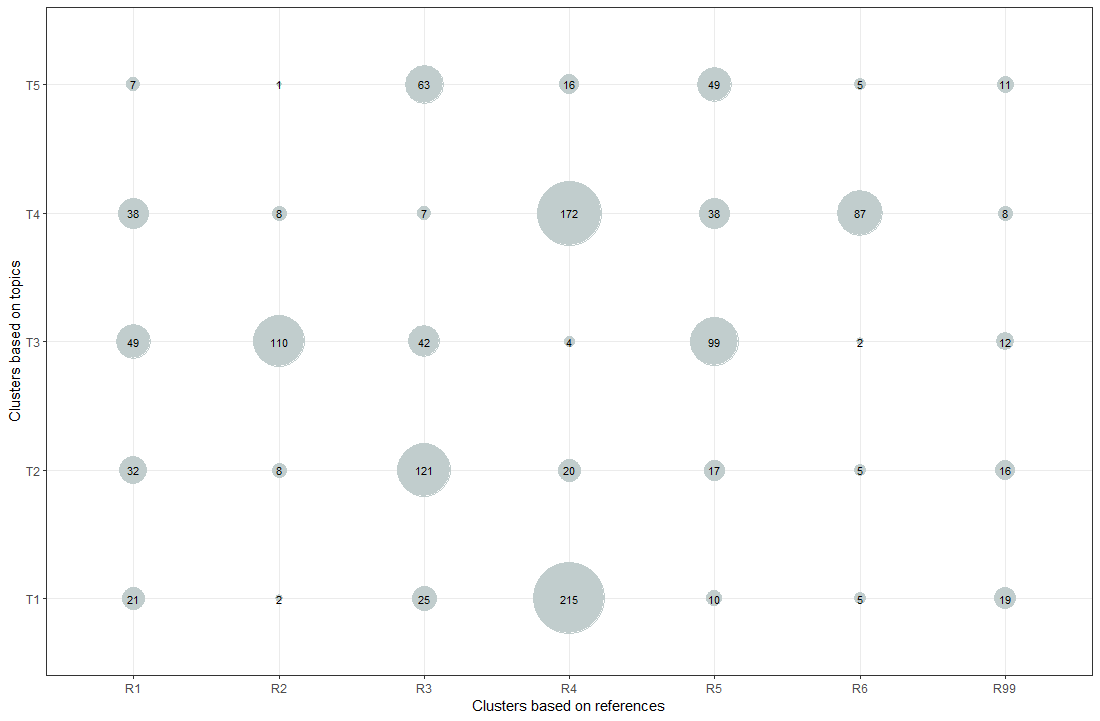} 
   \caption{Comparison of clusters in Topics\_20 and in cited references networks. The five clusters obtained through Louvain algorithm in the Topics\_20 network are in the $x$-axis and labelled as T1, \ldots , T5. The six big clusters obtained through Louvain algorithm in the cited reference are in the $y$-axis and labelled as R1, \ldots , R6; R99 collects 45 clusters with isolated articles or with a size less than 8 articles.}
   \label{fig:TvsR}
\end{sidewaysfigure}

More in detail, Figure \ref{fig:TvsR} compares the clusters obtained in Topic\_20 and cited-references networks. The Topic\_20 network is partitioned by Louvain algorithm in 5 clusters reported in rows and labelled T1, \ldots , T5. The cited-references network is instead partitioned in 51 clusters, 45 of which contain either isolated papers or groups of less than 8 articles. For the sake of simplicity Figure \ref{fig:TvsR} reports in columns the 6 largest clusters R1, \ldots , R6; plus a seventh one, labelled ``R99'', that contains the other 45 clusters. The largest 6 clusters collect 1,278 articles, i.e.\ the $95.1$\% of the total number of classified articles. 
The value of $0.645$ for Cramer's V (see Table \ref{cramer}) indicates a moderate association between the  classification based on Topic\_20 and the one based on cited references. Figure \ref{fig:TvsR} synthesizes the distinction between the two classifications: articles belonging to the same cluster on the basis of one information source are generally spread over at least three different clusters on the basis of the other. 

It is worth noting some meaningful overlapping. In particular, in absolute terms, the two most important overlaps are the one between R4 and T1 (215 papers, i.e. 50\% of R4 and 72\% of T1) and the one between R4 and T4 (172 papers, 40\% of R4 and 48\% of T4). Now, T1 contains papers mainly pertaining to the fields of industrial and labor economics, while T4 gathers papers focused on the history of economic thought and on economic methodology. In turn, R4 is characterized by the prevalence of papers in the institutionalist and evolutionary traditions, two approaches that in terms of topics are primarily devoted to the theory of the firm and of industrial organization and which, having their methodological premises as a distinctive mark, devote a great deal of space to methodological research. This explains the overlapping: R4 shares with T1 many works on industrial economics, and with T4 many papers with strong methodological content. Notice that cluster R4 has as most cited references works by Robert Nelson and Sidney Winter, Oliver Williamson, and Joseph A. Schumpeter that are clearly related to the institutionalist and evolutionary traditions. 
There are two other overlaps worth noting. First, the large majority of the articles in R6 are in T4. This is because R6 gathers methodological papers that are grouped in T4 with papers on the history of economic thought. Secondly, most of the papers belonging to R2 also belong to T3. R2 collects papers primarily connected to the Sraffian approach, which in T3 are combined with papers in the post-Keynesian tradition and with Marxist analyses so as to form a broader set that, following a now established practice, we can label ``Cambridge Economics'' \citep{marcuzzo}.

\begin{sidewaysfigure}
\centering
\includegraphics [scale=0.25]{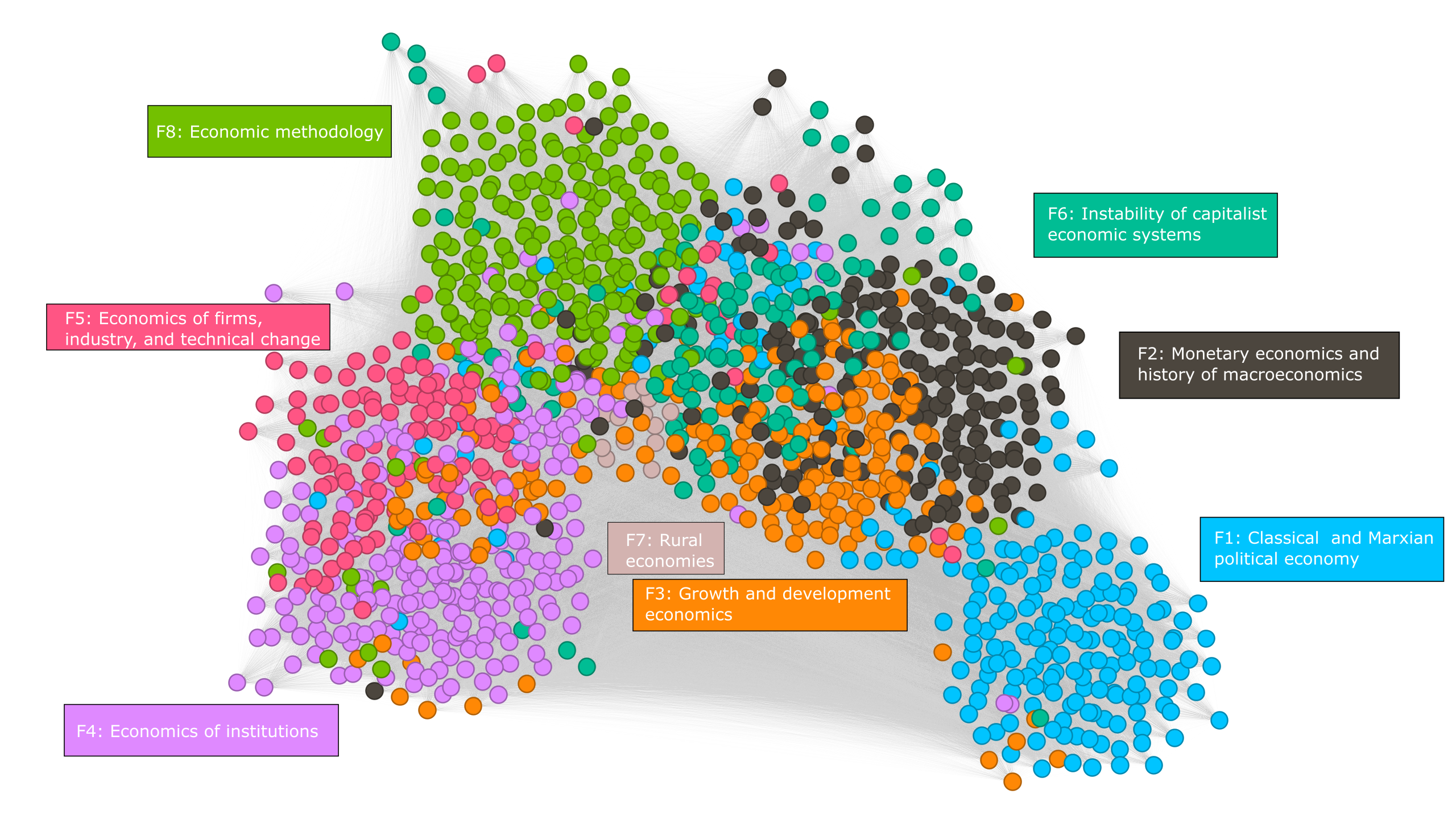} 
   \caption{The fused similarity network in the \textit{Cambridge Journal of Economics}. Nodes are articles. They are colored according to the clustering reported in Table \ref{cluster_colors}, obtained through the Louvain algorithm based on modularity in the Fused\_20 network. The graph is realized by \textsc{Gephi} with \code{OpenOrd} and \code{Noverlap} visualization algorithms.}
   \label{fig:net}
\end{sidewaysfigure}

Turning now to the analysis of the results obtained through the similarity network fusion, the Louvain algorithm detects 8 clusters in the Fused\_20 network. Table \ref{cluster_colors} reports the number and the proportion of articles classified in each group; group labels are described and justified in what follows; the color codes are the ones of Figure \ref{fig:net}.

\begin{sidewaystable}
\scriptsize
\caption{Clusters in the Fused\_20 network.}
    \centering
    \begin{tabular}{cllcc}
    \toprule
Cluster&Label&Color&N. of papers&\%\\
\hline
\definecolor{skyblue}{rgb}{0.53, 0.81, 0.92}
F1& \textsc{Classical and Marxian political economy}&\cellcolor{SkyBlue1}SkyBlue&210&15.6\\
F2&\textsc{Monetary economics and history of macroeconomics}&\cellcolor{black}\textcolor{white}{Black} &197&14.7\\
F3&\textsc{Growth and development economics}&\cellcolor{orange}Orange&189&14.1\\
F4&\textsc{Economics of institutions}&\cellcolor{Orchid2}Orchid2&246&18.3\\
F5&\textsc{Economics of firms, industry, and technical change}& \cellcolor{VioletRed2}VioletRed2&132&9.8\\
F6&\textsc{Instability of capitalist economic systems}& \cellcolor{Aquamarine3}Acquamarine3 &131&9.7\\
F7&\textsc{Rural economies}&\cellcolor{Bisque2}Bisque2&13&1.0\\
F8&\textsc{Economic methodology}&\cellcolor{PaleGreen3}PaleGreen3&226&16.8\\
\bottomrule
    \end{tabular}
    \label{cluster_colors}
\end{sidewaystable}

\begin{sidewaysfigure}
\centering
     \includegraphics [scale=0.5]{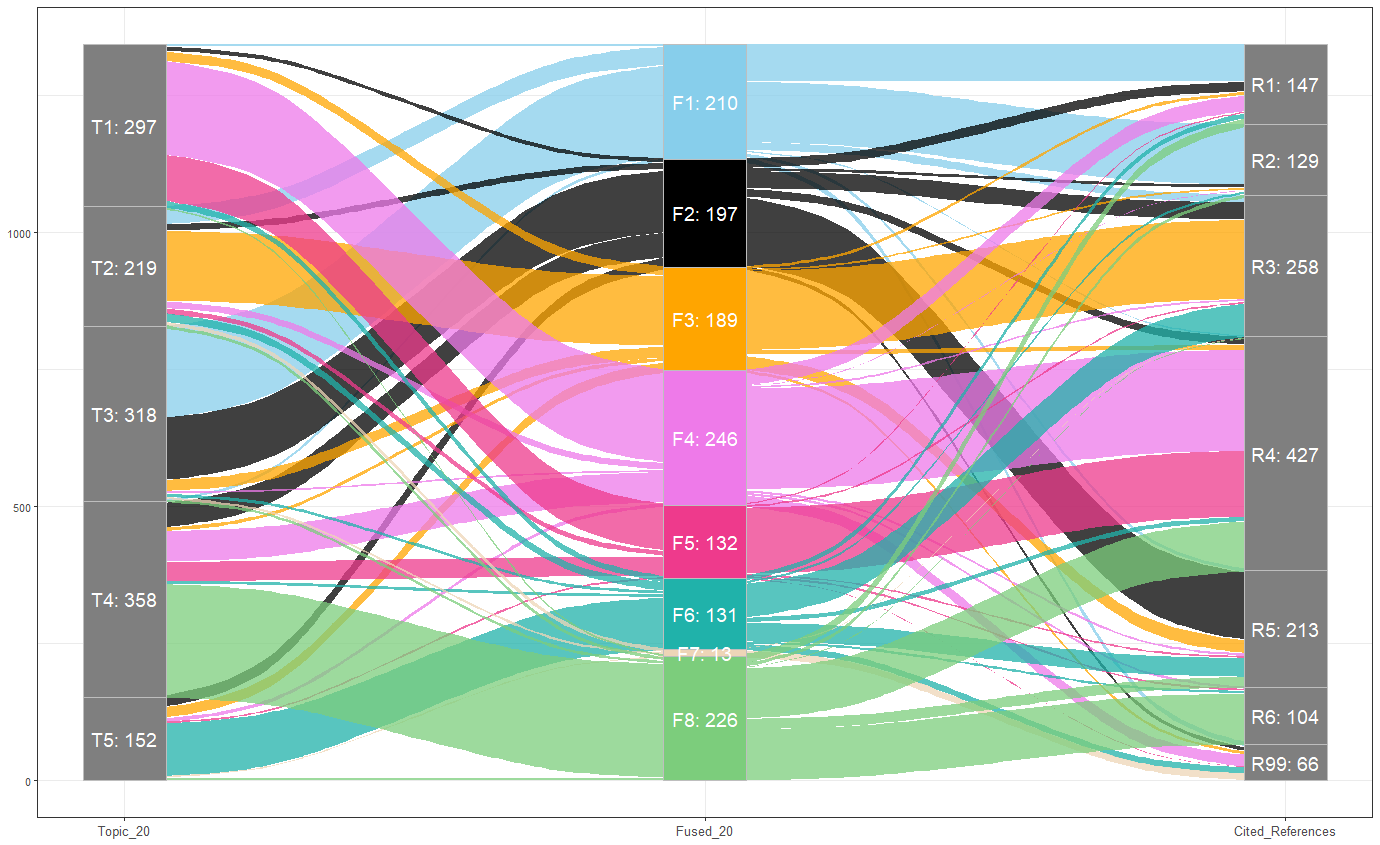} 
   \caption{Clusters of \textit{Cambridge Journal of Economics} articles in the fused network and in the networks based on Topic\_20 and Cited References. The strata of the alluvial plot refer to the clusters obtained in each network. The central stratum and the flows of the plot are colored as in Figure \ref{fig:net}. Strata report the codes and the number of articles of each cluster. The plot was realized in \proglang{R} environment by using the \pkg{ggalluvium} packages \cite{ggalluvial-package}.}
   \label{fig:alluvial}
\end{sidewaysfigure}

Figure \ref{fig:net} shows the fused network. The alluvial plot \citep{ggalluvial-article} of Figure \ref{fig:alluvial} compares the clusters obtained in the Fused\_20 network with the ones present in the Topic\_20 and Cited References network. The central stratum is composed of eight blocks corresponding to the clusters in the Fused\_20 network. The blocks are coloured according to Table \ref{cluster_colors}; their heights are proportional to the size of the clusters. The left stratum is composed of five blocks representing the clusters detected in the Topic\_20 network; the right stratum by seven blocks representing the clusters in the Cited Reference network. The flows between the central stratum and the other two lateral strata represent how the clusters in the central stratum are composed in terms of articles clustered in the two lateral strata. The flows are coloured according to the clusters in the Fused\_20 network, and the height of the flows is proportional to the size of the components contained in both blocks connected by the stream field. In the supplementary material Figure \ref{fig:crossclass} provides a different representation of the composition of clusters.

The first cluster F1 gathers 210 papers ($15.6\%$ of the articles) that primarily belong to the school of thought that can be defined as \textsc{Classical and Marxian Political Economy}, i.e., contributions in the tradition that was brought back to light by Piero Sraffa (1960) after it had been abandoned with the rise of the neoclassical paradigm.  The F1 cluster thus contains many papers on the history of economic thought dealing with the Classical economists and Marx. From the point of view of the topics addressed, it reflects the multiplicity of problems that formed the subject of these economists’ analyses, mainly the theory of value and distribution, fiscal policy, the theory of international trade, and the analysis of long-term trends of the economic system. The cluster also contains papers dealing with the critique of the neoclassical approach, thus expanding on the work undertaken in this respect by Sraffa, as well as contributions to the debate on Marxist theory and its relationship to Classical political economy. Marx’s \textit{Capital} figures as the second most cited document in the dataset, while the \textit{Wealth of Nations} by Adam Smith, who is arguably the father of Classical political economy, ranks sixth.
Figure \ref{fig:alluvial} shows that the \textsc{Classical and Marxian Political Economy} cluster brings together papers that the classification based on cited references places in two distinct groups: namely, contributions in the Marxist tradition gathered in cluster R1, and works in the vein of classical economists and Sraffa gathered in cluster R2. The cluster obtained from the fused network draws from two groups that the citation information keeps separate, pointing at a new, significant connection. The connection is all the more meaningful in that it is not clearly identifiable even by adopting textual information as a basis for classification. Indeed, while it is true that the papers in the cluster come mostly from T3, there is a significant proportion of papers coming from T2, mainly papers in the Marxist tradition dealing with macroeconomic issues. Most importantly, the papers from T3 account for only half of total T3, a cluster in which the Sraffian-oriented papers are grouped together with contributions on Keynesian and post-Keynesian theory belonging to the rather comprehensive group that we already called ``Cambridge economics''.

Hence, the F1 cluster is mainly characterized in terms of the theoretical approach. For the F2 and F3 clusters, instead, it is rather the homogeneity of the topics dealt with that justifies the grouping, although a certain prevalence of contributions in the post-Keynesian tradition can be observed. Both clusters collect contributions in the field of macroeconomics. 
The F2 cluster gathers 197 papers ($14.7\%$ of the articles) focused on \textsc{Monetary economics and the history of macroeconomics}. Analogously to the first cluster, also in this case the fusion emphasizes a significant connection which is less evident on the basis of the two classifications based on citation and textual information. Indeed, the \textsc{Monetary economics and the history of macroeconomics} cluster brings together papers that on the basis of cited references are distributed mainly between clusters R3 and R5, while on the basis of content, these papers are partly in T3 and partly in T4. The significance of the grouping we observe combining the two sources of information can be appreciated by taking into account the fact that the economist who is universally regarded as the father of macroeconomics, John Maynard Keynes, made one of his most innovative contributions in the theory of money. Now, most of the papers on the history of macroeconomics published in the \textit{Cambridge Journal of Economics} are devoted precisely to Keynes or to economists who draw on his contribution – post-Keynesian economists. These papers are thus largely concerned with monetary issues: in this regard, notice that Keynes’s \textit{General Theory} is the most cited document in the entire database. It seems therefore reasonable to group the contributions on monetary economics together with those on the history of macroeconomics.

The F3 cluster gathers 189 papers ($14.7\%$) mainly devoted to \textsc{Growth and development economics}. Most of the papers classified in this group are classified as T2 and R3. In this case, the fused network tends to reproduce the classifications obtained by considering topics and cited references separately. Indeed, most of the papers in R3 belong to T2, and most of T2 papers belong to R3.

The F4 cluster in the fused network, gathering 246 articles ($18.3\%$), is the largest and, at first sight, the most heterogeneous since it covers many research areas such as labor economics, comparative economic systems, social structure, law and economics. The label \textsc{Economics of institutions} indicates that all the areas of research represented in this cluster share a common thread in the analysis of the role of institutions in the economy.  This theme is explored in different directions, ranging from the consideration of the role of social institutions to the analysis of economic institutions and, with reference to the latter, investigating both single markets, such as the labor market, and the general structure of the economic system. The \textsc{Economics of institutions} cluster brings together almost half of the papers belonging to R4 with one-fifth of the papers belonging to R1. At the same time, it groups more than half of the papers in T1 with 15\% of the papers in T4.

As for the F5 cluster, this contains 132 papers ($9.8\%$) mainly dealing with the \textsc{Economics of firms, industry, and technical change}. From the point of view of the theoretical paradigm, there is a prevalence in this group of the evolutionary approach, a school of thought whose main contributions relate precisely to the theory of the firm, the economics of innovation and technical change, and the theory of industrial organization. In this case, too, the classification based on the whole set of information available isolates a set of papers that according to the two classifications based on topics and cited references do not form a group of their own. Indeed, while almost two-thirds of the papers in F5 cluster come from T1, these papers within T1 are found together with contributions on labor economics that are gathered in the fourth cluster in the fused network. Similarly, while almost all of the papers in the fifth cluster come from R4, these papers represent just over a quarter of R4, where they are collected together with contributions that are again gathered in the fourth cluster in the fused network.

The F6 cluster contains 131 papers ($9.7\%$) primarily concerned with the \textsc{Instability of capitalist economic systems}. The main issues addressed are related to financial markets, and financial fragility in particular, to the sustainability of fiscal policy, together with the large set of issues raised by the outbreak of the Great Recession, with a particular focus on the debate on the response to the crisis within the European Union. It is mainly the information coming from topics analysis that contributes to determining this cluster, with two-thirds of the papers coming from group T5; with respect to the classification based on cited references, this cluster mainly draws from two groups that collect macroeconomic papers, R3 and R5, in which however the subjects of instability and crises do not stand out distinctly.

The F7 cluster (labelled \textsc{Rural economies} ) represents a negligible component of the network, as it only contains 13 articles ($1.0\%$). It is worth mentioning, however, that this group has a clear characterization since all papers deal with topics related to rural economies, such as rural poverty, the role of informal credit markets, and agricultural production. These papers are classified into three different clusters according to topics and are almost all dispersed in R99.

Finally, the last F8 cluster gathers 226 articles ($16.8\%$) and can suitably be labelled \textsc{Economic methodology}. Quite reasonably, along with a large majority of articles with strong methodological content, in this group we find contributions that deal with the question of pluralism in economics or, more generally, with the sociology of economics. Once again, it is the textual information that provides the greatest contribution, as the cluster is mostly made up of articles from group T4; conversely, it draws from two distinct groups from the point of view of the classification based on cited references: indeed, it is true that there is a group with a clear methodological character, R6, but as we have seen an important share of methodological papers are found in R4, which gathers contributions in the institutionalist and evolutionary traditions.

Overall, it appears that the extreme heterogeneity that characterizes the papers published in the \textit{Cambridge Journal of Economics}, while making any attempt at classification more difficult, makes the similarity network fusion technique particularly valuable. The papers in our dataset lend themselves to being classified both on the basis of their analytical approach and on the basis of the topics covered, and the two classifications based on cited references and topics do not always favor the same criterion. It is therefore certainly interesting to combine textual and citation information in order to identify the criterion that yields the most strongly connected groups of papers.

\section{Conclusion}

The issue of using multiple sources of information for classifying papers and delineating research fields is an old problem in scientometrics. Usually, the delineation of scientific fields is conducted by considering only one layer of information at a time. There are classifications based on information about citations and references; and classifications that rely on content by adopting distant reading techniques. In most cases, the use of different sources of information results in different classifications of the same set of articles. 
So far, the attempts to use together multiple sources of information for classifying articles and delineating research fields have adopted techniques that require strong assumptions about statistical distributions of data, and about the weights to assign to different information when they are integrated. Moreover, they are usually limited to two layers of information. The Similarity Network Fusion technique proposed here is an unsupervised technique able to integrate more than two different layers of information in a single similarity network. SNF technique does not require any assumption about the statistical distribution of data, nor the choice of weights to be attributed to the layers of information when they are combined. 

The case-study addressed in this work regards the classification of articles published in the \textit{Cambridge Journal of Economics}. Founded in the 1970s, it is the leading generalist non-mainstream economics journal, open to contributions from diffent schools of economics. The task of classifying its articles is particularly difficult as they differ not only in terms of subjects but also in approaches to the same subject.

To this end, two layers of information are used: one based on bibliographic coupling, and the other based on contents. The bibliographic coupling served to define a similarity network among articles. The similarity of contents was defined by using two different approaches applied to the full-text of articles. The first one was based on Bags of Words, i.e. on the relative frequency of words in articles; the second one, requiring more statistical assumptions, was based on LDA topic modeling and produced six different similarity networks corresponding to different pre-defined numbers of topics.

The first result of the paper consists in showing that Bags of Words and LDA produce similarity networks highly associated. Hence, the use of one or the other technique, and the adoption of different numbers of topics do not entail a relevant loss of information. Indeed, when the same technique of clustering is applied to the seven networks based on contents, the resulting classifications of articles are highly associated.

The second result confirms previous analyses: the similarity network based on cited references and the ones based on contents have a moderate level of association; hence they convey different information about articles. Indeed, when the same technique of clustering is applied to the network based on cited references and to the networks based on contents, the resulting classifications of articles have a low level of association.

Thus, the adoption of a technique for integrating information about cited references and contents appears fully justified. 

The third result of the paper regards the application of the similarity network fusion technique. It results in seven different fused networks which have highly correlated structures, and which produced highly associated classifications of articles. A technique for showing the contribution of each layer to the structure of the fused network is also presented, and, in the present case, it shows that cited references contributed more to the final result than contents. 

The fourth result regards the comparison of SNF with other techniques for constructing hybrid similarity matrices by integrating different information. With respect to the integration obtained by using weighted means of original matrix, SNF has the main advantage of relieving researchers of the responsibility of choosing weights that can determine the final structure of the integrated network. SNF is based on local properties of the input networks, and the contribution of each starting network to the fusion can be rigorously measured \textit{ex-post}. Moreover, SNF can be used for integrating matrices whose similarities are calculated with different methods. This may determine very distant starting matrices for which a convex combination may be meaningless.
\citet{Baccini_2023} propose a new method for the integration of similarity matrices, and solve the problem of choosing weights in a very elegant way. Nevertheless, the method has very strict constraints about the spectral properties of the input similarity matrices (they are required to be ``completely positive"). Hence, it cannot be used for handling the data analysed in this paper. Therefore, for the present case study, SNF represents the best option. 

The classification obtained through SNF has been evaluated from the point of view of experts in the field, by inspecting whether it can be interpreted and labelled with reference to research programs and methodologies adopted in economics. Moreover, the classification obtained in the fused network is compared with the two classifications obtained when the networks of cited references and contents are treated separately. 

Overall, the classification obtained on the fused network appears to be fine-grained enough to represent the extreme heterogeneity characterizing contributions published in the \textit{Cambridge Journal of Economics}. The articles lend themselves to being classified both on the basis of their analytical approach and on the basis of the topics covered. The two classifications based on cited references and topics do not always favor the same criterion, thus resulting in less fine-tuned classifications.

The discussion of this last point has been conducted by considering the less favorable case, i.e. for the highest value of association between clusters obtained in the network based on cited references and the one based on topics. The discussion of results highlighted that the fine-grained classification obtained in the fused network appears qualitatively superior to the classifications obtained in the two layers separately. This suggests that fusion may be more effective in less challenging cases. In sum, the SNF technique appears as a useful tool for the complex task of fine-grained classification of articles.

The results presented here are promising but more research is needed. A first step might consist in comparing the classification obtained from SNF with the expert classification of articles defined by institutions or scholars. In the research agenda of this group, there is the comparison of the classification of articles obtained for the articles of the \textit{Cambridge Journal of Economics} with the classification (not so easily available) adopted by the \textit{Econlit} database, maintained by the American Economic Association.

A second line of research might consist in extending the analysis by adding other layers of information to the two adopted here, for instance by adding the similarity network based on co-citation among articles, on mentions of articles in social media, or on keywords chosen by the authors. These extensions would require a minimum adaptation of the setting presented in this work.  

A third line of research might consist in applying SNF technique to the classification of papers in a much larger scale. This will permit to explore the relation between granularity of classification and size of the set of papers to be classified. A first test may consist in verifying if the classification of CJE articles is stable in comparison with classification possibly obtained by considering a big set of  economics and social science journals, or a bigger multidisciplinary set of journals.

\section*{Declarations}

\begin{itemize}
\item Funding: the research is funded by the Italian Ministry of University, PRIN project: 2017MPXW98, PI: Alberto Baccini.
\item Conflict of interest/Competing interests: The authors have no competing interests to declare that are relevant to the content of this article.
\item Availability of data and materials: After acceptance, raw data will be available here \url{https://10.5281/zenodo.7876691}
\item Preprint: the article is available at \url{https://arxiv.org/pdf/2305.00026.pdf}
\item Authors' contributions: Alberto Baccini and Lucio Barabesi contributed to the study conception and design. Material preparation, data collection and analysis were performed by Alberto Baccini, Lucio Barabesi, Martina Cioni and Eugenio Petrovich; Federica Baccini supervised the methods of matrix integration and their comparison; Daria Pignalosa interpreted data from the methodology of economics perspective. All authors partecipated to the writing of the manuscript.
\end{itemize}

\bibliography{sn-bibliography}

\begin{appendices}
\section{Supplementary figures}

\counterwithin{figure}{section}

\begin{sidewaysfigure}
\centering
     \includegraphics [scale=0.50]{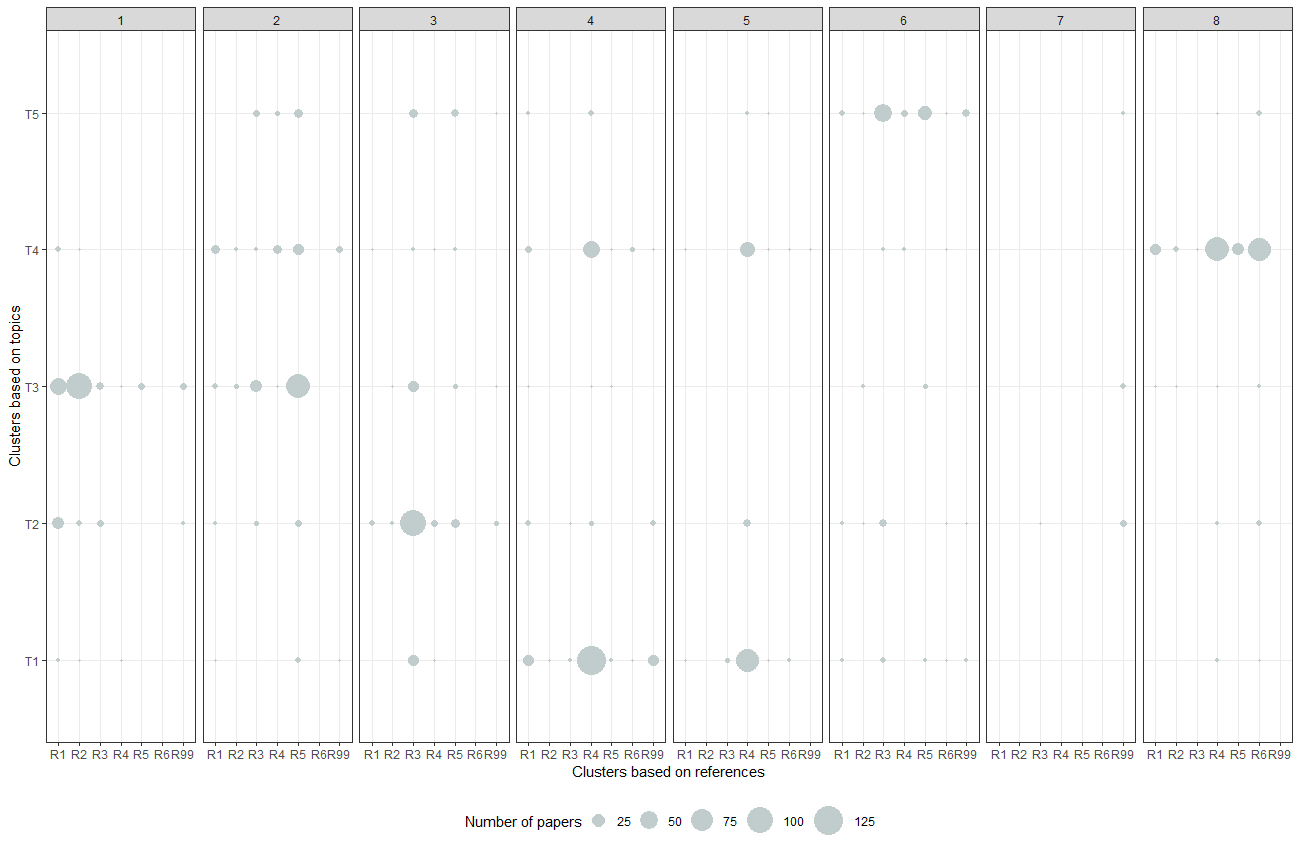} 
   \caption{Cross-distribution of articles from the \textit{Cambridge Journal of Economics} in different clusters. Each panel represents one of the 8 clusters obtained by Louvain algorithm applied to the Fused\_20 network. On the $y$-axis the 5 clusters obtained in the Topics\_20 network are reported; on the $x$-axis the clusters obtained in the Cited references network. Size of points is proportional to the number of papers. }
   \label{fig:crossclass}
\end{sidewaysfigure}
\end{appendices}

\end{document}